\documentclass[twocolumn,showpacs,preprintnumbers,amsmath,amssymb, superscriptaddress, prb]{revtex4-1}

\usepackage{multirow}
\usepackage{graphicx}
\usepackage{textcomp}
\usepackage{color}
\definecolor{red}{rgb}{1,0,0}
 
\definecolor{green}{rgb}{0,1,0}
 
\definecolor{blue}{rgb}{0,0,1}

\newcommand{\beq}{\begin{equation}}
\newcommand{\eneq}{\end{equation}} 
\newcommand{\beqa}{\begin{eqnarray}}
\newcommand{\eneqa}{\end{eqnarray}} 
\newcommand{\nn}{\nonumber} 
\newcommand{\bta}{\begin{tabular}}
\newcommand{\enta}{\end{tabular}}

\begin{document}

\title{Evolution of magnetic and crystal structures in the multiferroic FeTe$_2$O$_5$Br}

\author{M. Pregelj}
\affiliation{Jo\v{z}ef Stefan Institute, Jamova c.\ 39, SI-1000 Ljubljana, Slovenia}

\author{P. Jegli\v{c}}
\affiliation{Jo\v{z}ef Stefan Institute, Jamova c.\ 39, SI-1000 Ljubljana, Slovenia}
\affiliation{EN--FIST Centre of Excellence, Dunajska 156, SI-1000 Ljubljana, Slovenia}

\author{A. Zorko}
\affiliation{Jo\v{z}ef Stefan Institute, Jamova c.\ 39, SI-1000 Ljubljana, Slovenia}
\affiliation{EN--FIST Centre of Excellence, Dunajska 156, SI-1000 Ljubljana, Slovenia}

\author{O. Zaharko}
\affiliation{Laboratory for Neutron Scattering, PSI, CH-5232 Villigen, Switzerland}

\author{T. Apih}
\affiliation{Jo\v{z}ef Stefan Institute, Jamova c.\ 39, SI-1000 Ljubljana, Slovenia}
\affiliation{EN--FIST Centre of Excellence, Dunajska 156, SI-1000 Ljubljana, Slovenia}

\author{A. Gradi\v{s}ek}
\affiliation{Jo\v{z}ef Stefan Institute, Jamova c.\ 39, SI-1000 Ljubljana, Slovenia}

\author{M. Komelj}
\affiliation{Jo\v{z}ef Stefan Institute, Jamova c.\ 39, SI-1000 Ljubljana, Slovenia}

\author{H. Berger}
\affiliation{\'{E}cole Polytechnique F\'{e}d\'{e}rale de Lausanne, Switzerland}

\author{D. Ar\v{c}on}
\affiliation{Jo\v{z}ef Stefan Institute, Jamova c.\ 39, SI-1000 Ljubljana, Slovenia}
\affiliation{Faculty of mathematics and physics, University of Ljubljana, Jadranska c.\ 19, SI-1000 Ljubljana, Slovenia}

\date{\today}

\begin{abstract}

Neutron diffraction and nuclear quadrupole resonance (NQR) measurements were employed to investigate magnetic order in the non-ferroelectric phase preceding the low-temperature multiferroic state in FeTe$_2$O$_5$Br.
Refinement of the neutron diffraction data and simulations of $^{79,81}$Br NQR spectra reveal that the incommensurate magnetic ordering in the non-ferroelectric state comprises amplitude-modulated magnetic moments, similarly as in the multiferroic state.
The two ordered states differ in the orientation of the magnetic moments and phase shifts between modulation waves.
Surprisingly, all symmetry restrictions for the electric polarization are absent in both states.
The different ferroelectric responses of the two states are thus argued to arise from the differences in the phase shifts   between certain modulation waves, which cancel out in the non-ferrolectric state.

\end{abstract}


\maketitle

\section{Introduction}

The discovery of magnetically induced electric polarization\cite{Kimura, Kenzelmann05} 
revealed a new aspect and a great application potential of geometrically frustrated spin systems.\cite{ScottNature, KimuraARMR, Nmat07}
Frustration often leads to complex incommensurate (IC) magnetic structures, which can break the inversion symmetry and thus overcome a fundamental restriction for the macroscopic electric polarization.
Depending on the magnetic order, the magnetoelectric (ME) effect is associated with two types of exchange interaction.
In collinear spin structures the ME coupling is explained by changes of the isotropic exchange interaction leading to the exchange striction,\cite{SergienkoE, Kenzelmann05, Nmat07} whereas in spiral spin structures the antisymmetric part of the anisotropic exchange interaction is held responsible.\cite{SergienkoIDM, Katsura, Arima, Seki, Xiang}
Despite the generally accepted phenomenological description,\cite{ScottNature, Nmat07}
the microscopic picture of the ME mechanism is much more complex and still lacks a unified explanation.

In this paper we focus on the mechanism of the ME coupling in FeTe$_2$O$_5$Br,\cite{PregeljPRL, PregeljPRL2} where electric polarization in the long-range ordered elliptical IC amplitude-modulated (AMOD) magnetic state was suggested to originate from phase shifts between the exchange-coupled AMOD magnetic waves and thus argued to differ from conventional ME mechanisms.
The system adopts a layered structure of [Fe$_4$O$_{16}$]$^{20-}$ tetramer  clusters connected via Te$^{4+}$ ions.\cite{Becker} 
The magnetic lattice is composed of alternating antiferromagnetic Fe$^{3+}$ ($S$\,=\,5/2) spin chains coupled by frustrated interactions,\cite{PregeljAFMR} which amount to $\sim$\,1/3 of the dominant intrachain interaction $J_2$\,$\sim$\,19\,K.
Two subsequent magnetic transitions were identified.\cite{PregeljPRB}
The first, from paramagnetic to the high-temperature IC magnetic state (HT-IC) with a constant wave vector {\bf q}$_{\text{IC1}}$\,=\,($\frac{1}{2}$\,0.466\,0), occurs at $T_{N1}$\,=\,11\,K and is rapidly followed at $T_{N2}$\,=\,10.5\,K by the second one into the low-temperature IC multiferroic state (LT-IC).
The elliptical IC AMOD order in the LT-IC phase is characterized by long axis of the ellipsis along the (1,\,$-$1,\,\,0.2) direction in the $a^*bc$ orthonormal system (used throughout the paper),\cite{PregeljPRL2} and the magnetic wave vector that progressively changes from {\bf q}$_{\text{IC1}}$ to {\bf q}$_{\text{IC2}}$\,=\,($\frac{1}{2}$\,0.463\,0), where it settles below $T$\,$\sim$\,6\,K.\cite{PregeljPRB}
The accompanying electric polarization, ascribed to the exchange striction of the interchain interactions, points along the $c$-axis.\cite{PregeljPRL}
In contrast to the LT-IC state, the magnetic ordering in the HT-IC phase is still unknown and thus hampers the understanding why the electric polarization is absent in this phase and why it develops in the LT-IC phase.\cite{PregeljPRB}

Using combined nuclear magnetic and quadrupolar resonance (NMR and NQR, respectively), spherical neutron polarimetry (SNP), and neutron diffraction techniques, we solved the magnetic structure in the HT-IC phase.
Here the magnetic moments are, like in the LT-IC phase, sinusoidally modulated, but are now almost completely collinear with the $b$ axis.
Furthermore, our NQR results indicate changes of the electric-field gradient (EFG) at the Br sites below $T_{N2}$ (in the multiferroic LT-IC phase), corroborating the minute displacements of the Te$^{4+}$ ions that manifest as a bulk electric polarization.\cite{PregeljPRL} 
Comparison with the LT-IC magnetic structure implies that in the HT-IC state the phase shifts between certain magnetic AMOD waves are suppressed in accordance with the proposed ME coupling mechanism.\cite{PregeljPRL, PregeljAFMR}
These phase shifts are thus most likely responsible for the lack of the electric polarization above $T_{N2}$.

\section{Experimental Details}

NMR and NQR measurements were performed on high-quality single crystals \cite{PregeljPRL} with an average size of $15\times8\times2$\,mm$^3$ on a home-build spectrometer in the temperature range between 4 and 300\,K in zero magnetic field and at 4.7 and 9.4\,T.

Spherical neutron polarimetry (SNP) was performed at 10.7\,K on the same crystals using a MuPAD device on the triple axis spectrometer TASP ($\lambda$\,=\,3.2\,\AA) at the Swiss Neutron Spallation Source (SINQ), Paul Scherrer Institute (PSI), Switzerland. Intensities of the magnetic reflections at temperatures between 9.2 and 11\,K were collected at the same location using the single-crystal diffractometer TriCS ($\lambda$\,=\,2.32\,\AA).

\section{Results}

\subsection{NQR and NMR experiments}

Local-probe NQR and NMR experiments on $^{79}$Br and $^{81}$Br nuclei with the spin $I$\,=\,3/2 were chosen because, (i) in addition to a standard dipolar magnetic moment through which they detect the local magnetism, (ii) they also possess a quadrupole moment, making them sensitive to EFG and thus highly susceptible even to the tiniest structural deformations. 
From the experimental point of view, however, this makes it very difficult to find the resonant frequency, since EFG can vary within several orders of magnitude, depending on the details of the local Br environment.
In addition, in FeTe$_2$O$_5$Br there are two crystallographically inequivalent Br sites; Br$_1$ that is coupled to a single magnetic Fe$^{3+}$ ($S$\,=\,5/2) ion, and Br$_2$, interacting with three magnetic Fe$^{3+}$ ions [inset to Fig.\,\ref{NMR}(b)].

\subsubsection{Characterization of local Br environment}

\begin{figure} [!]
\includegraphics[width=0.50\textwidth]{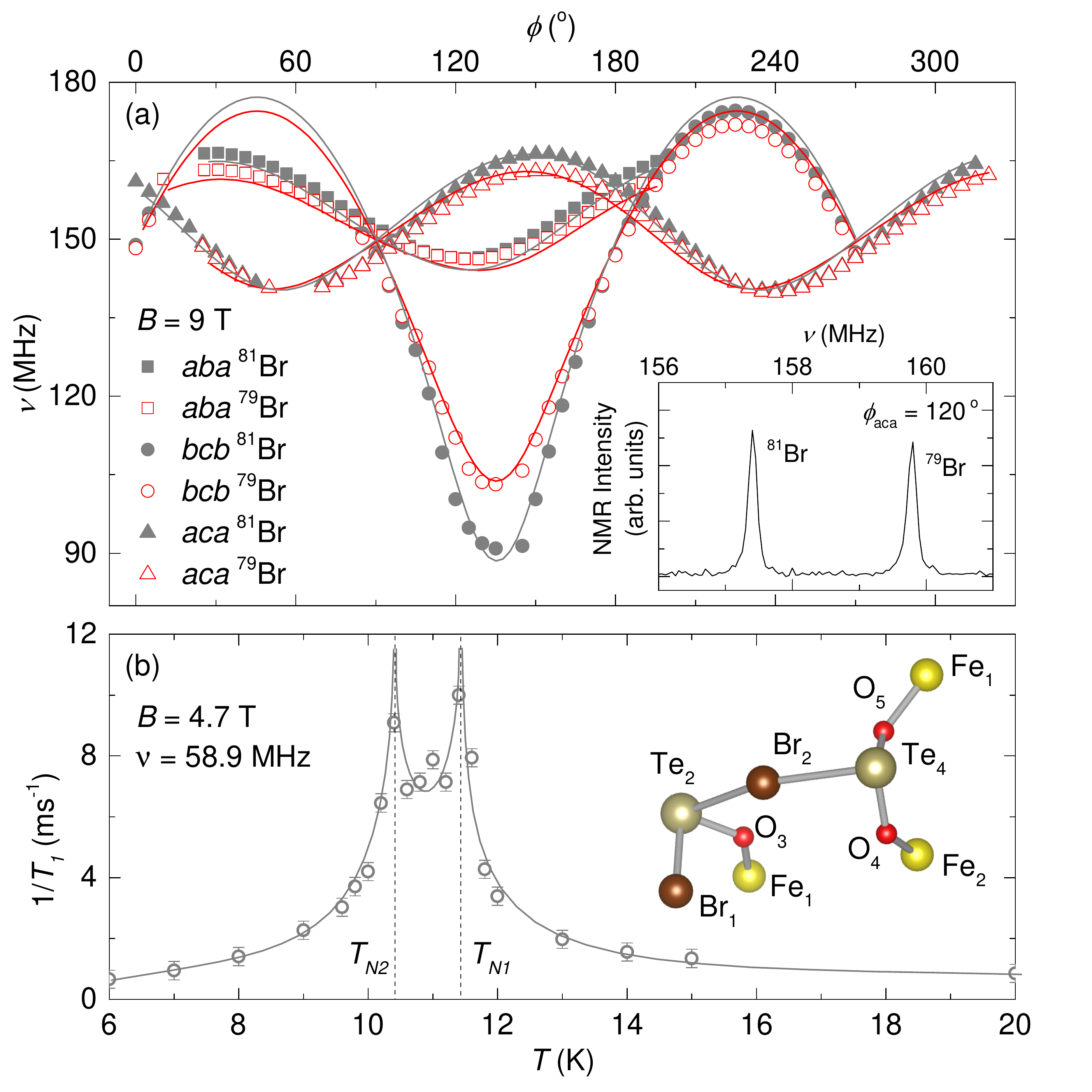}
\caption{(Color online) (a) Angular dependences of the $^{81}$Br and the $^{79}$Br NMR central lines  ($-1/2\leftrightarrow1/2$) at 80\,K and 9\,T (symbols) for the Br$_1$ site and simulations based on DFT calculated EFG tensor (lines). Inset: The corresponding NMR spectrum of both isotopes at $\phi_{\text{aca}}$\,=\,120\,$^\circ$.
(b) Temperature dependence of the NMR relaxation rate 1/$T_1$ at the peak of the central $^{81}$Br$_2$ line, i.e., at 58.9\,MHz, in the field of 4.7\,T along $a^*$. Solid lines are guides for the eyes. Inset: local coordinations of the Br$_1$ and Br$_2$ sites.}
\label{NMR}
\end{figure}

\begin{table} [!]
\caption{Upper panel: DFT calculated components of the electric-field gradient (EFG) tensor $V_{ij}$ given for the Br$_1$ and Br$_2$ sites in the $a^*bc$ coordinate system in units of 10$^{21}$\,V/m$^2$. Lower panel: the corresponding quadrupole splitting $^{79,81}\nu_Q$\,=\,$\frac{6\,^{79,81}QeV_{zz}}{4I(2I-1) h}$ and axial asymmetry parameters $\eta$\,=\,$(V_{xx}$\,$-$\,$V_{yy})/V_{zz}$.
\label{tabEFG}}
\begin{ruledtabular}
\begin{tabular}{ccccccc}
  \multicolumn{3}{c} {Br$_1$} && \multicolumn{3}{c} {Br$_2$} \\
\hline
  -15.8353 & -13.9572 & -14.8939 &&  2.96904 & -0.47202 & -2.45593  \\
  -13.9572 & 7.00619  & 34.5642 && -0.47202  & 3.35194 & 10.6256  \\
  -14.8939 & 34.5642  & 8.82908 && -2.45593 & 10.6256 & -6.32098 \\
\hline
\multirow{2}{*} {$\eta_Q$=0.09} &  \multicolumn{2}{c} {$^{79}\nu_Q$=184.88 MHz} && \multirow{2}{*} {$\eta_Q$=0.58} & \multicolumn{2}{c} {$^{79}\nu_Q$=50.49 MHz} \\
&  \multicolumn{2}{c} {$^{81}\nu_Q$=154.76 MHz} && & \multicolumn{2}{c} {$^{81}\nu_Q$=42.26 MHz} \\
\end{tabular}
\end{ruledtabular}
\end{table}
We first performed density functional theory (DFT) calculations of the EFG tensors at the Br$_1$ and Br$_2$ sites (for details see Appendix\,\ref{appB}), in order to facilitate the search of the $^{79}$Br and $^{81}$Br NQR and NMR signals.
The obtained EFG tensors (Table\,\ref{tabEFG}) imply that the quadrupole splitting $^{79,81}\nu_Q$\,=\,$\frac{6\,^{79,81}QeV_{zz}}{4I(2I-1) h}$ at the Br$_1$ site is significantly larger than at the Br$_2$ site.
Here, $^{79,81}Q$ denotes the quadrupole moment of the $^{79,81}$Br isotopes, $h$ is the Planck's constant, $e$ is the electron charge, while $V_{ij}$ ($i,j$\,=\,$x,y,z$) are the components of the EFG tensor.
The EFG tensor for Br$_1$ is almost axially symmetric with the asymmetry parameter $\eta$\,=\,$(V_{xx}-V_{yy})/V_{zz}$\,=\,0.09, whereas for the Br$_2$ site $\eta$\,=\,0.58.
The $^{81}$Br NQR signals at 260\,K (deep in the paramagnetic phase) were experimentally found at 165.0 and 38.4\,MHz for Br$_1$ and Br$_2$, respectively  (Fig.\,\ref{figNQR-T-dep}), which is within $\sim$10\,\% of the values\cite{Abragam} $\nu_Q$(1+$\eta^2$/3)$^\frac{1}{2}$ predicted by DFT calculations.
The remarkable accuracy of the DFT results was further tested by measuring angular dependences of the $^{79,81}$Br NMR signals for Br$_1$ around all three crystallographic axes ($a^*$, $b$, and $c$) in the field of 9\,T at 80\,K [Fig.\,\ref{NMR}(a)].
If Br$_1$ EFG values are increased by 8.2\,\% with respect to the DFT calculations, the angular dependence, calculated by exact diagonalization of the nuclear spin Hamiltonian (for details see Appendix\,\ref{appA}), nicely matches the experiment [lines in Fig.\,\ref{NMR}(a)].
We note that small discrepancy between calculations and experimental data probably originates from additional hyperfine fields and/or tiny misalignment of the crystal.

Having determined the quadrupolar interactions, it is now our task to clarify if $^{81}$Br NMR probes the magnetism as well.
We measured spin lattice relaxation 1/$T_1$, which is a highly sensitive parameter for the critical spin fluctuations in the vicinity of the magnetic transitions.
The temperature dependence of 1/$T_1$ for the $^{81}$Br$_2$ central NMR transition [Fig.\,\ref{NMR}(b)], at 58.9\,MHz, clearly reflects two distinct lambda-type anomalies, signifying the two magnetic transitions.
Compared to the zero field results the splitting between the two transitions increases by $\sim$0.5\,K, due to the magnetic field of 4.7\,T applied along the $a^*$ axis.\cite{PregeljPRB}

\subsubsection{Low temperature NQR spectra}

\begin{figure} [!]
\includegraphics[width=0.50\textwidth]{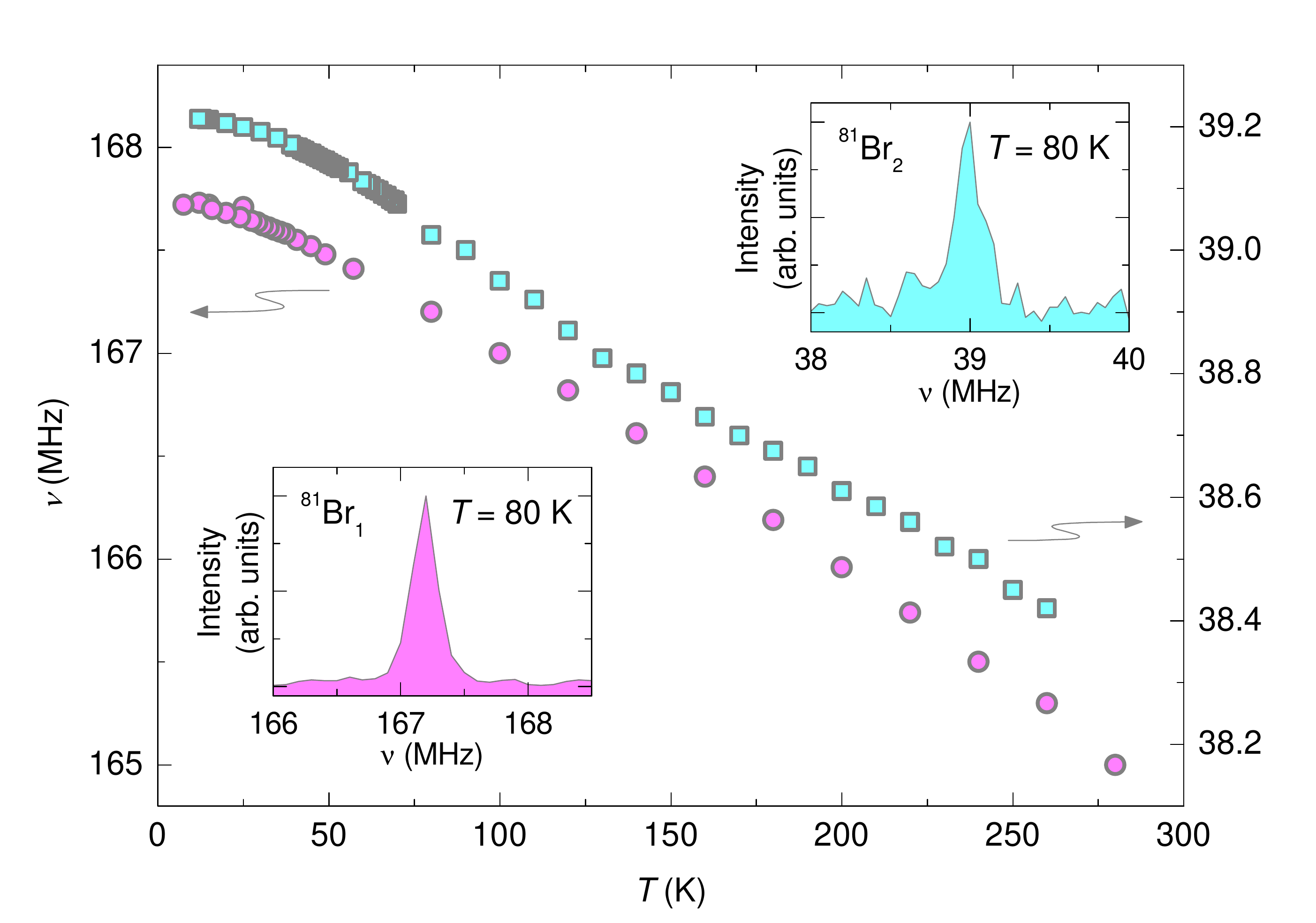}
\caption{Temperature dependence of the $^{81}$Br NQR signal for the Br$_1$ and the Br$_2$ sites. Insets: Corresponding spectra measured at 80\,K.}
\label{figNQR-T-dep}
\end{figure}
\begin{figure} [!]
\includegraphics[width=0.50\textwidth]{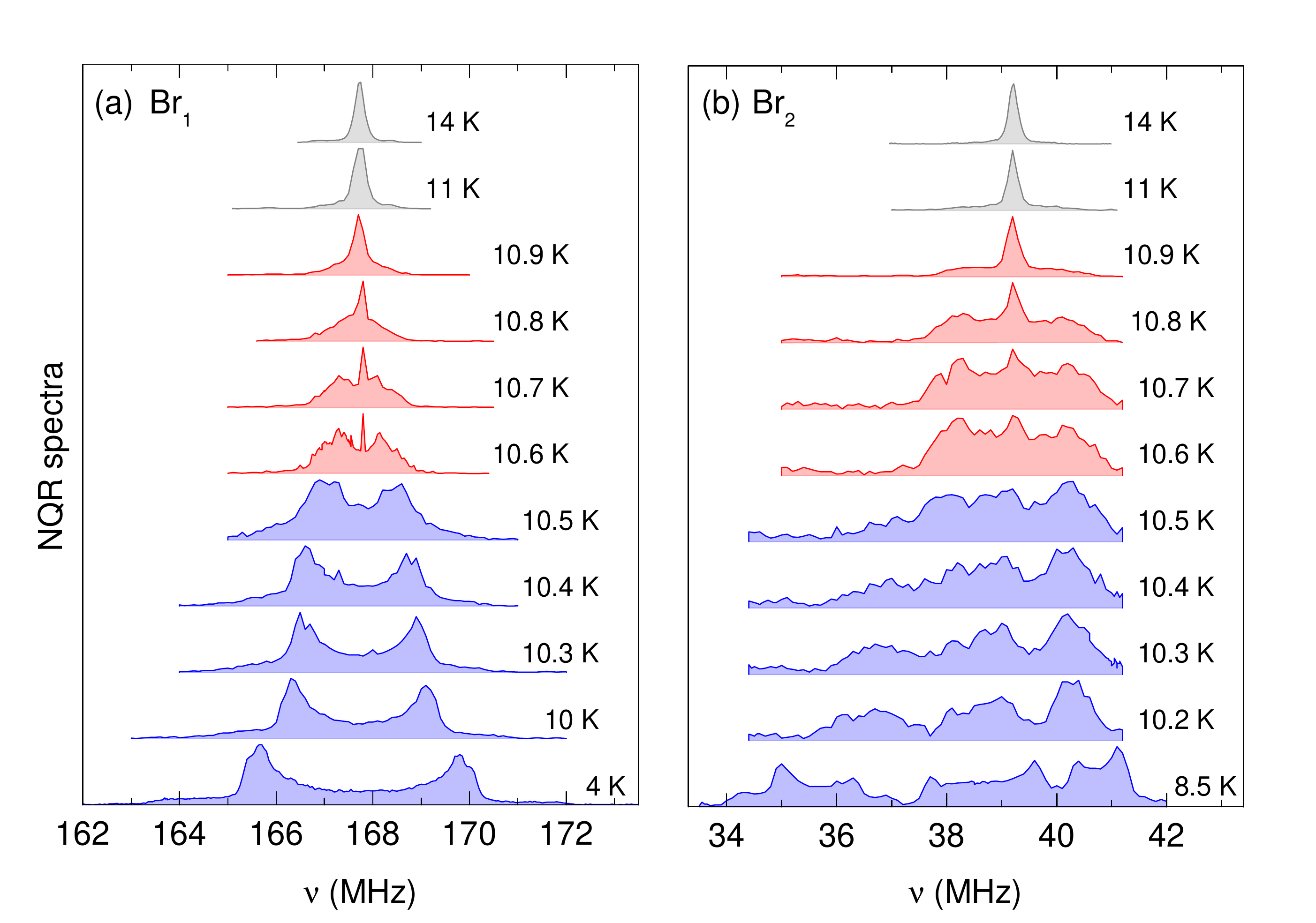}
\caption{(Color online) Normalized $^{81}$Br NQR spectra for (a) the Br$_1$ and (b) the Br$_2$ sites. Different colors (shades) correspond to the three different magnetic phases (paramagnetic, HT-IC and LT-IC).}
\label{figNQRspec}
\end{figure}
\begin{figure} [!]
\includegraphics[width=0.50\textwidth]{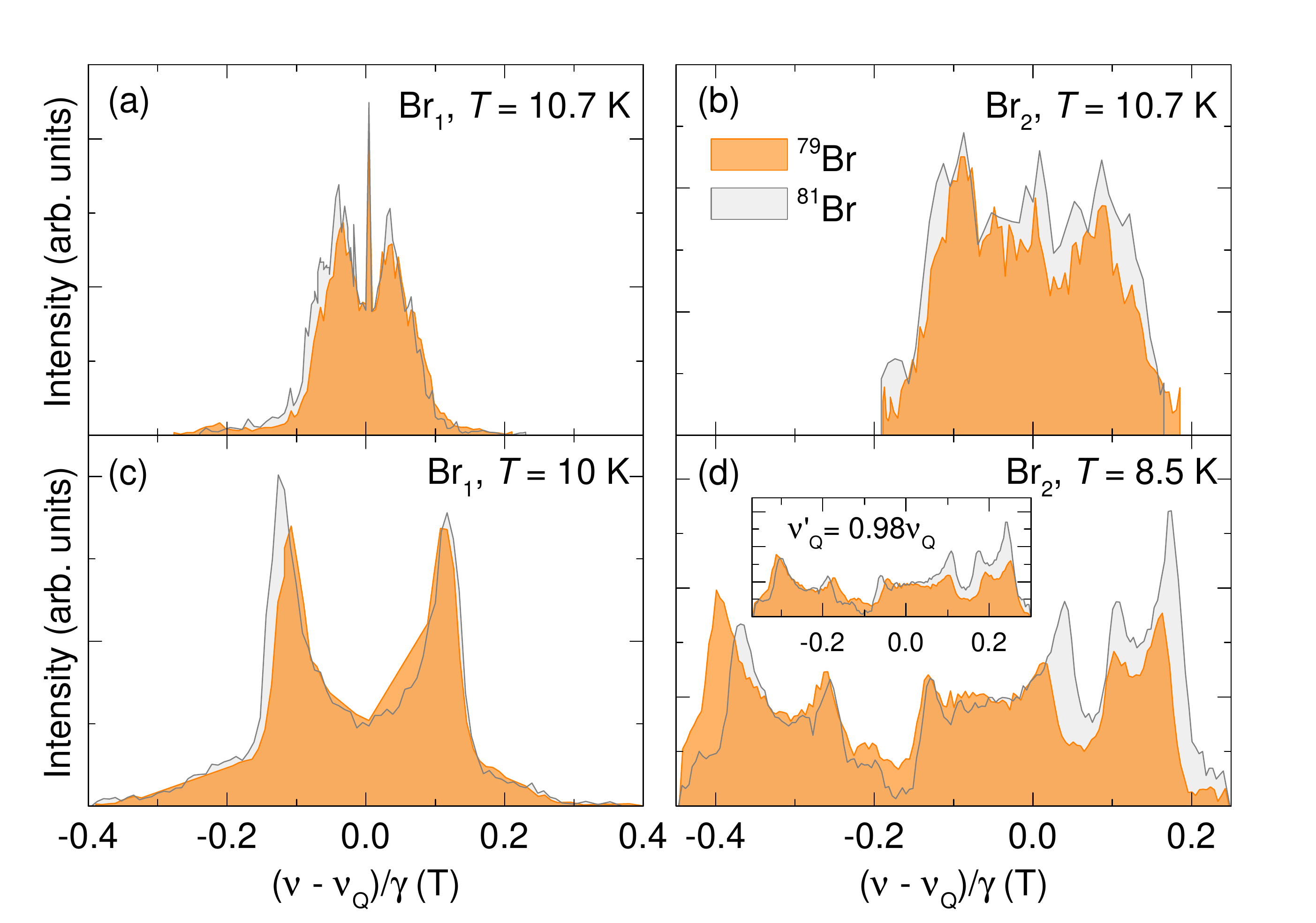}
\caption{(Color online) The $^{81}$Br and $^{79}$Br NQR spectra for (a) the Br$_1$ and (b) the Br$_2$ sites in the HT-IC phase and for (c) the Br$_1$ and (d) the Br$_2$ sites in the LT-IC phase. In the inset a scaling with reduced $\nu_Q^\prime$\,=\,0.98$\nu_Q$ is shown for Br$_2$.}
\label{NQRhtltcomp}
\end{figure}

On cooling from 260\,K both $^{81}$Br NQR lines (Br$_1$ and Br$_2$) shift linearly to higher frequencies down to $\sim$30\,K, where they are no longer temperature dependent (Fig.\,\ref{figNQR-T-dep}).
This shift is a result of a slight ($\sim$2\,\%) increase of the EFG values due to the crystal lattice contraction.
As expected, below the first magnetic transition, at $T_{N1}$\,=\,11\,K, the intensity of the sharp paramagnetic resonance gradually transfers to a broad U-shaped signal (Fig.\,\ref{figNQRspec}), typical for a sinusoidal distribution of local magnetic fields in the IC structures.\cite{Blinc} 
Clearly, the two signals coexist in a narrow temperature region around $T_{N1}$, i.e., approximately between 11.0 and 10.6\,K, revealing the first-order nature of this transition.
To clarify the origin of the IC modulation in the HT-IC phase, i.e., whether is it solely magnetic or also structural, we in parallel measured NQR signals for the $^{79}$Br isotope, which has smaller gyromagnetic ratio $\gamma$ and larger quarupolar moment $Q$ than the $^{81}$Br isotope ($^{79}\gamma$\,=\,10.6663\,MHz/T~$<$~$^{81}\gamma$\,=\,11.4978\,MHz/T and $^{79}Q$\,=\,31.3$\times$10$^{-30}$\,m$^2$~$>$~$^{81}Q$\,=\,26.2$\times$10$^{-30}$\,m$^2$). 
Overplotting the $^{79}$Br and the $^{81}$Br signals in the HT-IC phase we find that for both sites their widths scale with $\gamma$'s [Figs.\,\ref{NQRhtltcomp}(a),(b)] and not with $Q$'s.
This proves that the observed IC modulation is solely magnetic.
In addition, the center of gravity of the lines does not shift at $T_{N1}$ (Fig.\,\ref{figNQRspec}), indicating that the EFG tensors are unaltered, i.e., suggesting that magnetic transition at $T_{N1}$ does not induce any significant crystal structure distortions.

On further cooling, both signals dramatically change again at $T_{N2}$\,=\,10.5\,K, i.e., at the transition from the HT-IC to the LT-IC phase.
Clearly, to achieve proper scaling of the $^{79}$Br and $^{81}$Br LT-IC signals with $\gamma$'s [Fig.\,\ref{NQRhtltcomp}(d)], $\nu_Q$ at the Br$_2$ site has to be reduced by $\sim$2\,\% [inset to Fig.\,\ref{NQRhtltcomp}(d)].
This implies that EFG at the Br$_2$ site is sensitive to minute lattice distortions, accompanying the electric polarization in the LT-IC phase.\cite{PregeljPRL}
Modification of $\nu_Q$ for the Br$_1$ site is to small to be assessed from our measurements [Fig.\,\ref{NQRhtltcomp}(c)].
In addition, as opposed to the HT-IC phase where both Br$_1$ and Br$_2$ sites have simple U-shaped NQR spectra, in the LT-IC phase their spectra suddenly become completely different (Fig.\,\ref{figNQRspec}).
Further splitting of the Br$_2$ spectra below $T_{N2}$ implies the loss of certain symmetries in the multiferroic state that might be still present in the HT-IC phase.

\subsection{Neutron diffraction experiments}

\subsubsection{Magnetic structure in the HT-IC phase}

\begin{figure*} [!]
\includegraphics[width=1.0\textwidth]{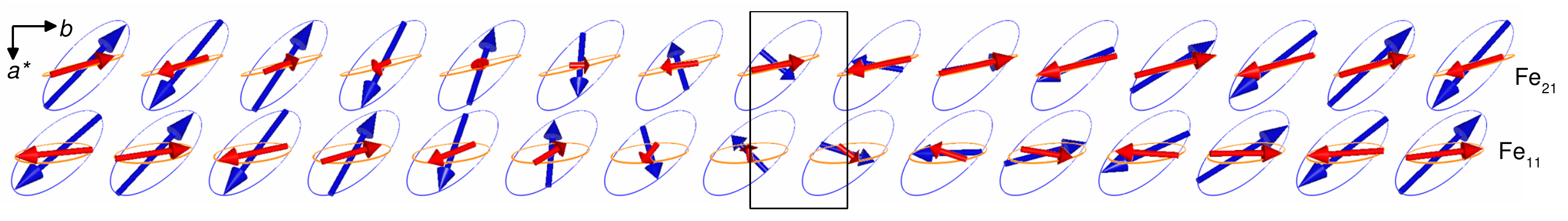}
\caption{(Color online) Evolution of Fe$_{11}$ and Fe$_{21}$ magnetic moments in the $a^*b$ projection along the $b$ axis in the HT-IC [light (red) arrows] and LT-IC [dark (blue) arrows] phases. The rectangle represents the unit cell. Note that the elliptical envelop for the HT-IC phase is almost completely flattened, with moments pointing approximately  along the $b$ axis.}
\label{figMagHT}
\end{figure*}

To verify whether the different NQR spectra originate from different magnetic-order symmetries in the HT-IC and the LT-IC phases, we next decided for neutron diffraction experiments, aiming to determine
the magnetic order in the HT phase. The combination of spherical neutron polarimetry (SNP) and conventional single crystal neutron diffraction has proven very useful in the past for determination of complex magnetic structures, e.g., IC arrangements or systems with superimposed nuclear and magnetic contributions.\cite{NeutronScatt}
Compared to conventional single-crystal neutron diffraction, the SNP method has enhanced sensitivity to the direction of the magnetic moments and thus allows to differentiate between complex magnetic structures, e.g., between AMOD and helical spin arrangements. 
The two experiments were conducted at 10.7\,K, with SNP performed for three different crystal orientations. In addition to the $hk0$ orientation, where the scattering plane was defined by the (1\,0\,0) and (0\,1\,0) reciprocal vectors, the crystal was rotated to the scattering plane defined by the (0\,1\,0) and (1\,0\,2) vectors, and finally to the scattering plane with (0\,0\,1) and either the (0.5\;0.466\;0) or (0.5\;0.534\;0) vectors. Altogether, we accumulated 24 polarization matrices and 62 integrated intensities.

Starting with the representation analysis, we find that magnetic wave vector {\bf q}$_{\text{IC1}}$\,=\,($\frac{1}{2}$\,0.466\,0) breaks the inversion symmetry already in the HT-IC phase.
This leaves two possible one-dimensional irreducible representations of the little (magnetic) group, which couple magnetic moments at the Fe sites related by a 2$_{1y}$ twofold screw axis.\cite{PregeljPRL}
Since the presence of the 2$_{1y}$ symmetry would explain the lack of the electric polarization (in the $ac$ plane) as well as the high symmetry of the NQR spectra in the HT-IC phase, we start the refinement of the corresponding magnetic order assuming a single irreducible representation.
In the most general case, the magnetic moment at a particular Fe site is defined as
\begin{equation}
{\bf{S}}_{mn}({\bf{r}}_{i})  =  {\bf{S}}_{0\,mn}^{\text{Re}}\cos({\bf{q}}\cdot{\bf{r}}_{i} - \psi_{mn}) 
 +  {\bf{S}}_{0\,mn}^{\text{Im}}\sin( {\bf{q}}\cdot{\bf{r}}_{i} - \psi_{mn}). 
\end{equation} 
Here, the vector ${\bf{r}}_i$ defines the origin of the $i$-th unit cell, $m$\,=\,1,2 identifies the crystallographically inequivalent Fe-sites, and $n$=1-4 denotes the four Fe positions within the crystallographic unit cell (for details see the caption of Table\,\ref{tab1}).
The complex vector ${\bf{S}}_{0\,mn}$ is determined by its real and imaginary components, ${\bf{S}}_{0\,mn}^{\text{Re}}$ and ${\bf{S}}_{0\,mn}^{\text{Im}}$, which define the amplitude and the orientation of the magnetic moments, i.e., the envelope of the magnetic cycloid/spiral, while $\psi_{mn}$ denotes its phase shift. 
The magnetic wave vector {\bf q} is in units of (2$\pi/a$,\,2$\pi/b$,\,2$\pi/c$) and $\psi_{mn}$ in 2$\pi$.
We stress that within a single irreducible representation ${\bf{S}}_{0\,m(n+2)}$\,=\,($\pm$1,\,$\mp$1,\,$\pm$1)\,$\cdot$\,${\bf{S}}_{0\,mn}$ and $\psi_{m(n+2)}$\,=\,$\psi_{mn}$\,+\,q$_{IC1}^y$/2, where $n$\,=\,1,\,2.
To avoid overparameterization of the problem we assume the same complex vector ${\bf{S}}_{0\,m1}$\,=\,${\bf{S}}_{0\,m2}$\,$\equiv$\,${\bf{S}}_{0\,m}$.
Moreover, to assure the best assessment of the experimental uncertainty, the estimated standard deviations of the polarization matrices and the overall refinement were treated in the same way as in our study of the LT-IC phase.\cite{PregeljPRL2}
Surprisingly, neither of the two irreducible representations can describe the HT-IC data satisfactory, as both refinements diverge, implying that all symmetry operations are broken already in the HT-IC phase.

\begin{table} [!]
\caption{Components of vectors {\bf S}$_{0\,m}^s$\,=\,($S_{0\,x}^s$,\,$S_{0\,y}^s$,\,$S_{0\,x}^s$) for $s$\,=\,Re,\,Im, defining the elliptical envelops for two independent magnetic atoms (Fe$_1$ and Fe$_2$) for the best magnetic structure model at 10.7\,K, and eight magnetic phases $\psi_{mn}$ in units of 2$\pi$, i.e., one for each of the magnetic Fe$_{mn}$ atoms in the unit cell ($m$\,=\,1,2, $n$\,=\,1-4). The sites Fe$_{12}$-Fe$_{14}$ are obtained from Fe$_{11}$ [$0.1184(6)$, $-0.001(1)$, $0.9734(7)$] and Fe$_{22}$-Fe$_{24}$ from Fe$_{21}$ [$0.9377(6)$, $0.2953(1)$, $0.8562(6)$] by symmetry elements $i$, $2_{1y}$ and $2_{1y}i$, respectively. The orientation of the moments is given in the $a^*bc$ coordinate system, while $|{\bf S}_{0}|$\,$\approx$\,1.2\,$\mu_B$.
\label{tab1}}
\begin{ruledtabular}
\begin{tabular}{ccccc}
     $s$ = Re, Im   & Fe$_1^{\text{Re}}$ & Fe$_1^{\text{Im}}$ & Fe$_2^{\text{Re}}$ & Fe$_2^{\text{Im}}$   \\
\hline
$S_{0\,x}^s$/$|{\bf S}_{0\,m}^s|$        & 0.98   & 0.10   &   0.71 &   0.23       \\
$S_{0\,y}^s$/$|{\bf S}_{0\,m}^s|$        & 0.14   &-0.97   &   0.35 &   0.94       \\
$S_{0\,z}^s$/$|{\bf S}_{0\,m}^s|$        & 0.16   & 0.24   &   0.61 &   0.26       \\ 
\hline
$|{\bf S}_{0\,m}^s|/|{\bf S}_{0}|$  & 0.21   & 0.92   &  0.07  &  1.00  \\
\hline
\hline
 $m$ &  $\psi_{m1}$ & $\psi_{m2}$ & $\psi_{m3}$ & $\psi_{m4}$\\
\hline
 1     &  0.00            &  0.02          &    0.32         &     0.39 \\
 2     &  0.89            &  0.86          &    0.01         &     0.01  \\
\end{tabular}
\end{ruledtabular}
\end{table}

In the next step we, therefore, resort to the elliptical IC structure model used to describe the LT-IC phase.
Here, the symmetry relations between ${\bf{S}}_{0\,m(n+2)}$ and ${\bf{S}}_{0\,mn}$ and between $\psi_{m(n+2)}$ and $\psi_{mn}$ for $n$\,=\,1,2 do not exist anymore, while we extend the relation ${\bf{S}}_{0\,mn}$\,$\equiv$\,${\bf{S}}_{0\,m}$ to include $n$\,=\,3,4.
In addition, as in the LT-IC case,\cite{PregeljPRL2} we allow different domain populations for each experiment.
Indeed, this refinement leads to a stable solution, which is almost completely sinusoidally modulated with  {\bf S}$_{0\,m}^{\text{Im}}$\,$\gg$\,{\bf S}$_{0\,m}^{\text{Re}}$ and the dominant components ({\bf S}$_{0\,m}^{\text{Im}}$) of the magnetic moments aligned very close to the $b$ axis (Fig.\,\ref{figMagHT}, Table\,\ref{tab1}).
Goodness of the refinement reflects in the total cost $C_{\text{tot}}$\,=$\,\sum_j \chi_j^2$/($N_{j\,\text{obs}}$--$N_{\text{par}}$)=\,46.6,
which is close to $C_{\text{tot}}$\,=\,33.7 obtained in the refinement of the LT-IC structure\cite{PregeljPRL2} and better than $C_{\text{tot}}$\,=\,57.3 for a simplified sinusoidal (collinear) AMOD model of the HT-IC phase.
It is significantly better than  $C_{\text{tot}}$\,=\,169.3 for an alternative circular cycloidal model.
In the above expression for $C_{\text{tot}}$ $j$ is the number of the datasets and $N_{\text{par}}$ is the number of the fitting parameters.
For each dataset with $N_{{j\,\text{obs}}}$ observations $\chi_j^2$\,=\,$\sum_{i=1}^{N_{j\,\text{obs}}}(X_{i\,\text{obs}}$\,--\,$X_{i\,\text{calc}})^2/\sigma_{X_{i\,\text{obs}}}^2$,  where $X$ denotes polarization matrix elements $P$ or integrated intensities $I$, and $\sigma_{X_{i\,\text{obs}}}$ is the estimated standard deviation of the observation.
We point out that the refinement of the LT-IC phase included larger $N_{{j\,\text{obs}}}$, while $N_{\text{par}}$ was the same, which led to a somewhat lower $C_{\text{tot}}$.
In short, the results of our refinement surprisingly show that magnetic ordering removes all symmetry restrictions for the electric polarization already in the HT-IC phase.

\subsubsection{Temperature dependence}

In order to understand why there is no electric polarization in the HT-IC phase, we next measured temperature dependence of 20 magnetic reflections between 11.0\,K and 9.2\,K.
This allows us to follow the evolution of the magnetic structure during the transition from the HT-IC to the LT-IC phase.
\begin{figure} [b!]
\includegraphics[width=0.50\textwidth]{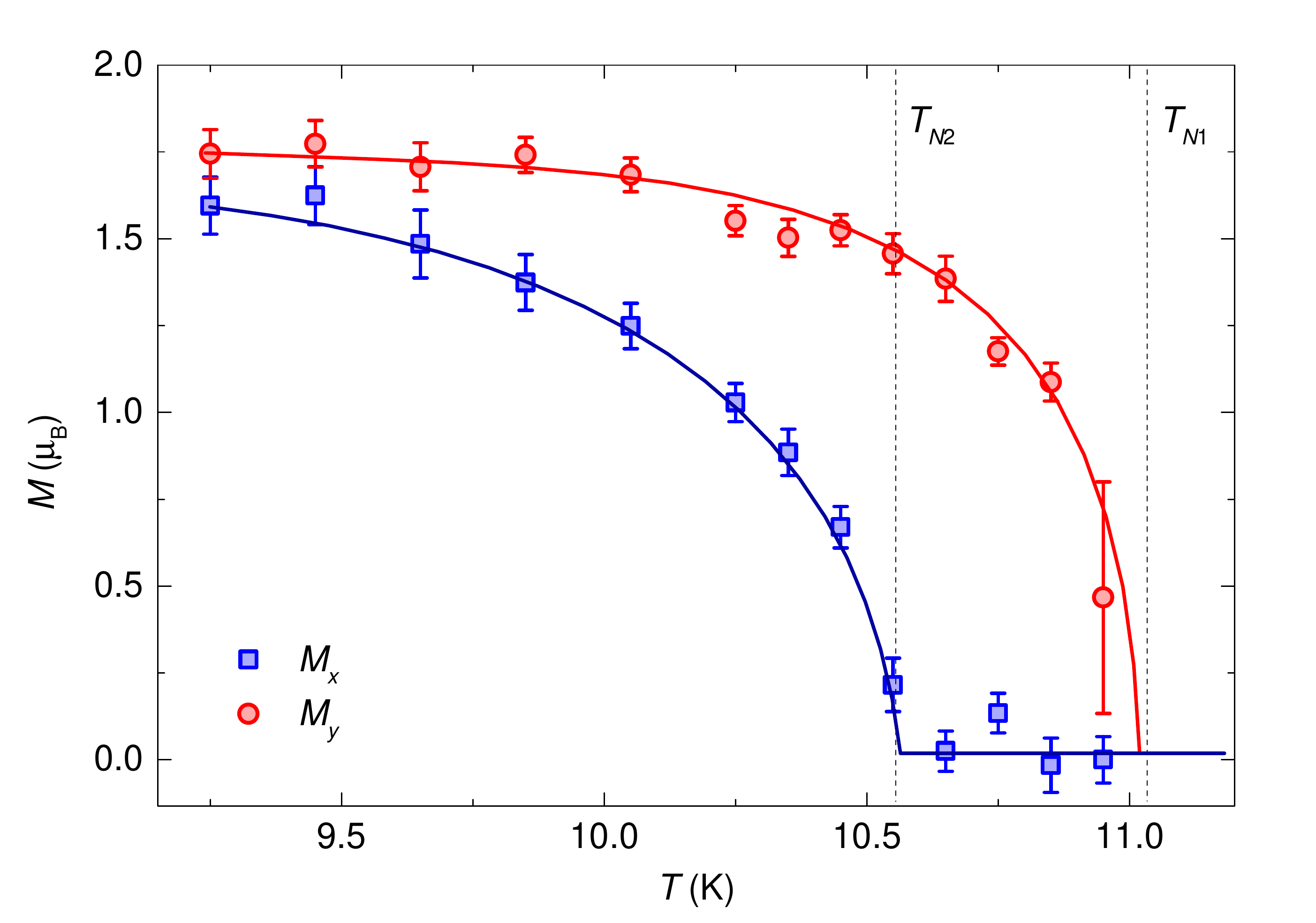}
\caption{(Color online) Temperature dependence of the $M_x$ and $M_y$ components of the magnetic moments deduced from the intensities of the neutron reflections on heating through the transition from the LT-IC to the HT-IC phase. Solid lines are guides for the eyes.}
\label{figMagPar}
\end{figure}
Due to the limited amount of data, rather small ordered magnetic moments in the investigated temperature interval and since the simpler collinear AMOD model already captures the most essential properties of the more complicated elliptical model our refinement considers the collinear sinusoidal AMOD magnetic structure model.
In particular, we assume that magnetic moments lie in the $a^*b$ plane and are strictly AMOD with the same ${\bf S}_0$\,$\equiv$\,${\bf S}_0^{\text{Im}}$ for all Fe sites.
This way we focus on magnetic phase shifts $\psi_{mn}$, which were pointed out as potential source of the ME coupling in earlier studies.\cite{PregeljPRL}
The results clearly show that with increasing temperature the $a^*$ component ($M_x$) is reduced and completely disappears at $T_{N2}$, whereas the $b$ component ($M_y$) starts to decrease only in the vicinity of $T_{N1}$ (Fig.\,\ref{figMagPar}).
This indicates that the Fe$^{3+}$ magnetic moments turn from the (1\,-1\,0) direction in the LT-IC phase towards the (0\,1\,0) direction in the HT-IC phase.
On the other hand, the changes of $\psi_{mn}$ are minute and appear to be insensitive to the LT-IC to HT-IC transition. 
In fact, reducing the number of independent $\psi_{mn}$ only mildly affects the quality of the refinement, i.e., $R$ factor increases from $\sim$6 to $\sim$9, and it does not affect the derived rotation of the magnetic moments.
This suggests that our temperature-dependent-neutron diffraction data are insufficient to reliably extract the temperature evolution of the magnetic phases.
Similarly, we were unable to refine the temperature dependence of the tiny magnetic component along $c$  ($M_z$).

\section{Discussion}

\subsection{Long-range magnetic ordering}

At first sight, the results of the two complementary experimental techniques are contradictory. The neutron diffraction suggests that magnetic ordering breaks all crystal-symmetry relations already in the HT-IC phase, whereas NQR implies that the HT-IC phase is more symmetric than the LT-IC phase. 
In order to clarify this issue and to extract as much information about the long-range ordering in FeTe$_2$O$_5$Br as possible, hyperfine coupling tensors have to be determined.
Since both Br sites (Br$_1$ and Br$_2$) lie at general positions and since the hyperfine coupling interaction is symmetric in the first order, we need to find for each hyperfine coupling tensor all six components.
For simplicity we start with Br$_1$, which is coupled to a single Fe ion and thus its hyperfine interaction depends only on one hyperfine coupling tensor.
Taking into account the LT-IC magnetic structure\cite{PregeljPRL2} and the obtained EFG tensor (Figs.\,\ref{NMR},\,\ref{figNQR-T-dep}), we determine the Br$_1$ hyperfine coupling tensor by fitting the NQR Br$_1$ spectrum measured at 4\,K (for details of the NQR spectrum calculations see Appendix\,\ref{appA}).
We stress that the derivation of the hyperfine coupling tensor from the angular dependences of the paramagnetic $^{79,81}$Br NMR signal [Fig.\,\ref{NMR}(a)] has been avoided due to possible crystal misalignments ($\leq$5\,$^\circ$), which can for so large EFG's result in resonance shifts of several MHz.
\begin{figure} [!]
\includegraphics[width=0.50\textwidth]{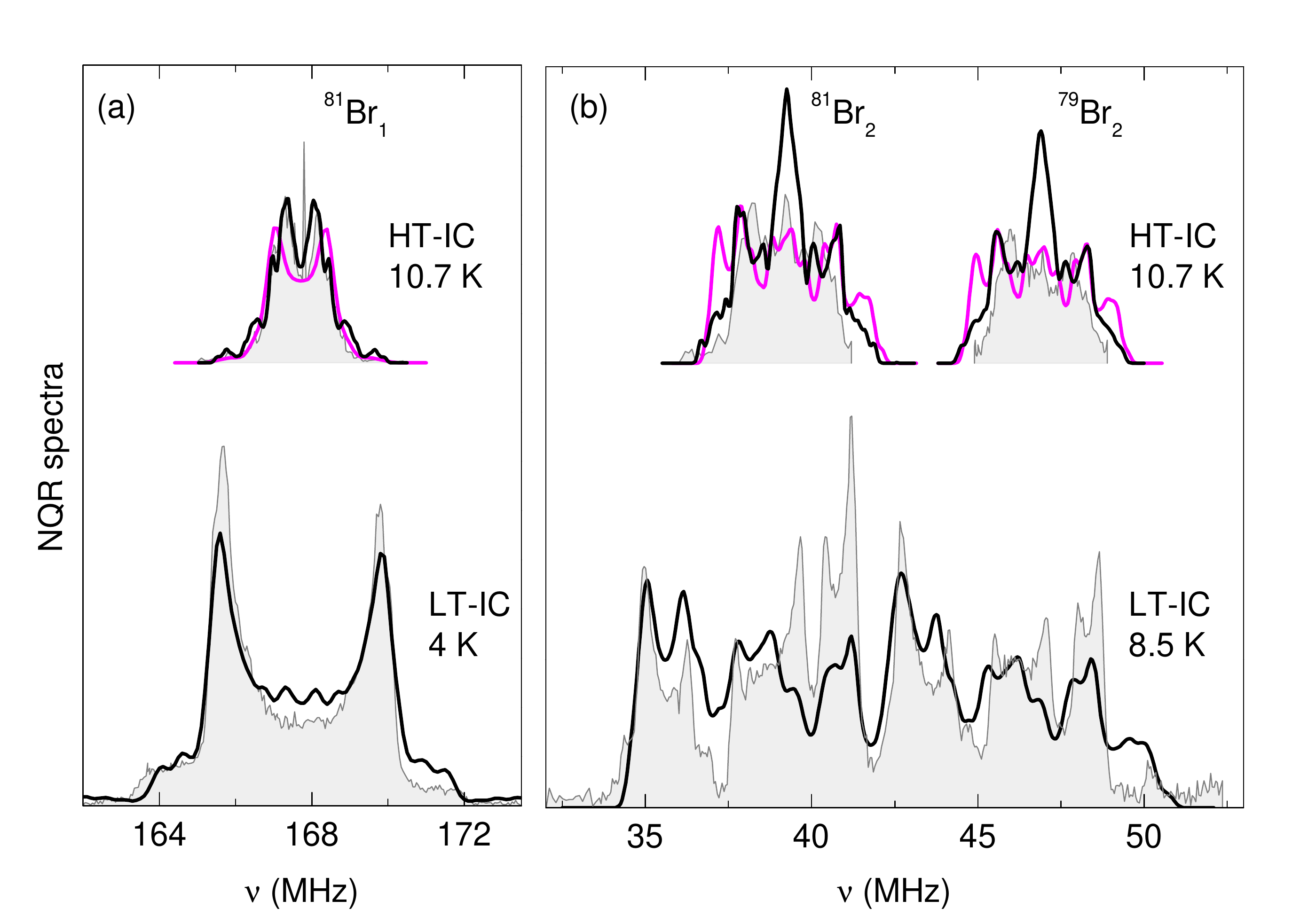}
\caption{(Color online) The $^{81}$Br NQR spectra for (a) Br$_1$ and (b) for both isotopes of Br$_2$ in the HT-IC and the LT-IC phases. Thick black lines are simulations considering magnetic structures determined by neutron diffraction and derived hyperfine coupling tensors (see text for details). For comparison we show simulations of the HT-IC spectra based on the LT-IC structure with appropriately reduced size of the magnetic moments [thick light (magenta) line]. }
\label{figNQRsim}
\end{figure}
The obtained Br$_1$ hyperfine coupling tensor (Table\,\ref{tabHYP}) yields a good agreement between the experimental and calculated spectra [bottom panel in Fig.\,\ref{figNQRsim}(a)]. 
Moreover, considering the HT-IC magnetic structure (Table\,\ref{tab1}) and the derived hyperfine coupling tensor (Table\,\ref{tabHYP}), the  HT-IC $^{81}$Br$_1$ NQR spectrum is reproduced with a high accuracy [top panel in Fig.\,\ref{figNQRsim}(a)] with no adjustable parameters.
This unambiguously validates the orientation and amplitude modulation of the Fe$^{3+}$ moments in the HT-IC magnetic structure, as determined by neutron diffraction.
Quite importantly, if the small $c$ ($M_z$) component of the magnetic moments is neglected, the HT-IC spectrum  cannot be reproduced satisfactory, whereas disregarding the $M_x$ component (along the $a^*$) or ellipticity has almost no effect on the simulated spectrum.
\begin{table} [!]
\caption{Derived hyperfine coupling tensors for the Br$_1$ and Br$_2$ sites in the $a^*bc$ coordinate system in units of mT/$\mu_B$.
\label{tabHYP}}
\begin{ruledtabular}
\begin{tabular}{ccccccc}
\multicolumn{3}{c} {Br$_{1}$(Te$_2$-O$_3$-Fe$_1$)}  && \multicolumn{3}{c} {Br$_{2}$(Te$_4$-O$_5$-Fe$_1$)}\\
\hline
  -34.4 & -18.5 & -16.3 &&  128.7  &  -0.7  &   35.7 \\
  -18.5 &  53.0 &  -3.3 &&   -0.7  & -17.8  &   26.8 \\
  -16.3 &  -3.3 & 138.3 &&   35.7  &  26.8  & -226.0 \\
\hline
\multicolumn{3}{c} {Br$_{2}$(Te$_2$-O$_3$-Fe$_1$)} && \multicolumn{3}{c} {Br$_{2}$(Te$_4$-O$_4$-Fe$_2$)} \\
\hline
 147.0  &    28.9 &   40.1 && -116.7  &     5.5 &  -24.5 \\
  28.9  &    13.5 &   -5.3 &&    5.5  &   -14.3 &  -13.0 \\
  40.1  &    -5.3 & -258.3 &&  -24.5  &   -13.0 &   23.9 \\
\end{tabular}
\end{ruledtabular}
\end{table}

Encouraged by a very good agreement between the neutron scattering results and NQR data for the Br$_1$ site, we focus now on the more complicated Br$_2$ NQR spectrum.
Applying the same procedure, i.e., fitting of the LT-IC NQR spectrum by adjusting the hyperfine coupling tensor (the EFG is reduced by 14\,\% compared to DFT calculations) and considering the coupling with three different Fe$^{3+}$ sites, each with its own hyperfine-coupling tensor, we manage to reproduce the main features of the Br$_2$ NQR spectrum as well [bottom panel in Fig.\,\ref{figNQRsim}(b)].
To ensure a proper scaling of the EFG tensor, spectra for both isotopes were fitted simultaneously.
Our fitting results get even greater value when one considers that during such a broad frequency sweep the measured NQR intensity may significantly vary due to frequency-dependent sensitivity of the spectrometer and the use of several experimental setups.
In addition, we note that the discrepancy between the experimental and the calculated spectrum can be also due to tiny modulation of the $\nu_Q$, which can result from the weak IC structural  modulation found in SNP study of the LT-IC phase.\cite{PregeljPRL2}
This further complication is beyond the scope of our simulations, but may explain why the $^{79}$Br$_2$ and $^{81}$Br$_2$ NQR spectra scaled by $\gamma$'s do not match perfectly even when reduced $\nu_Q$ in the LT-IC phase is considered [inset in Fig.\,\ref{NQRhtltcomp}(b)].
Nevertheless, the derived hyperfine coupling tensors (Table\,\ref{tabHYP}) can reproduce the width and the main spectral singularities also for the HT-IC phase [black lines in the top panel in Fig.\,\ref{figNQRsim}(b)] when the EFG is increased by 2\,\% compared to the LT-IC phase, in accordance with our previous observations [Figs.\,\ref{NQRhtltcomp}(b),(d)].

For comparison, we plot also calculated spectra corresponding to the LT-IC magnetic structure with appropriately scaled magnetic moments to match the HT-IC values.
For these the discrepancy from the experimental data is much more pronounced for both Br sites [light lines in Fig.\,\ref{figNQRsim}].
The overall agreement between the neutron diffraction and the NQR results thus offers a confirmation of the proposed magnetic structures in the LT-IC and the HT-IC phases.
It also shows that the simplicity of the Br$_2$ NQR spectra in the HT-IC phase originates from a smaller size of the ordered magnetic moments and a changed orientation of spins, and not from the higher symmetry of the magnetic structure.
Finally, we note that the similarity between $\psi_{mn}$ and $\psi_{m(n+1)}$ for $n$=1,3, i.e., between the sites, related by the inversion symmetry, implies that even though the inversion symmetry is broken already by {\bf q}$_{\text{IC1}}$, the system effectively reduces this effect by matching the relevant phases.

\subsection{Magnetoelectric coupling}

The most important experimental finding of our study is that all crystal-symmetry relations are broken already by the HT-IC magnetic order, which should thus, in principle, also allow for the establishment of the electric polarization.
This is in line with the first-order nature of the transition, suggested by the coexistence of the HT-IC and the paramagnetic phase in a narrow temperature range  around $T_{N1}$ (Fig.\,\ref{figNQRspec}). 
Surprisingly, the electric polarization does not develop until the second magnetic transition into the LT-IC phase.
In addition, the lack of the symmetry relations removes all limitations regarding its orientation, leaving no clue why the actual polarization points along the $c$ axis.
The observed response, therefore, deviates from other multiferroics where the multiferroic phase evolves in two subsequent [or one as in RbFe(MoO$_4$)$_2$]\cite{RFMO} continuous magnetic transitions and thus allows to exploit the phenomenological description of the ME coupling to its full extent.\cite{ScottNature,Nmat07}
In particular, in contrast to our case, continuous transitions preserve certain relations between the magnetic ordering in the multiferroic phase and the symmetries of the crystallographic space group, which enables predictions of the direction of the emergent electric polarization.\cite{Harris}

In FeTe$_2$O$_5$Br, the electric polarization could, in principle, be associated with the reorientation of the magnetic moments from the (1\,-1\,0) direction in the LT-IC phase towards the (0\,1\,0) direction in the HT-IC phase.
However, such scenario would be most probably associated with the simultaneous reorientation of the electric polarization, which contradicts our previous results (Ref.\,\onlinecite{PregeljPRL}) showing that the orientation of the electric polarization is temperature independent in the entire LT-IC phase. 
A second possibility is that electric polarization is associated with the ellipticity in the LT-IC magnetic structure, which is again not very likely, as our refinement shows finite ellipticity also in the HT-IC phase.

To understand the absence of the electric polarization in the HT-IC phase, we thus compare the LT-IC and HT-IC magnetic structures in respect to the so-called {\em magnetic-phase-shift} ME coupling mechanism, which predicts electric polarization $P$\,$\propto$\,$M_i\cdot M_j\sin(\Delta\psi_k)$.\cite{PregeljPRL}
Here $M_{i,j}$ are the magnetic order parameters corresponding to a pair of the exchange-coupled AMOD magnetic waves, while $\Delta\psi_{k}$ denotes the phase shift between the two. 
Generally, each of the six possible exchange interaction $J_k$ ($k$\,=\,1-6)\cite{PregeljAFMR} can be involved in the ME coupling mechanism.
We find, however, that on heating from the LT-IC to the HT-IC state $\Delta\psi_{k}$ changes towards $\pi$ or to 0, i.e., leading to $P$\,$\to$\,0, only for $k$\,=\,4. 
Similar, but significantly weaker, trend is noticed for $J_5$, which has been also highlighted before as the most likely candidate to drive the exchange striction.\cite{PregeljAFMR} 
On the other hand, all changes corresponding to other $k$'s are rather irregular. 
These observations are further supported by the fact that the NQR spectrum for Br$_2$, which is coupled to the $J_4$-bridging Te$_4$ ion, shows a pronounced change of $\nu_Q$ and thus reveals tiny structural transformations, most likely related to the onset of the electric polarization.

The above argumentation, therefore, suggests that the most coherent explanation of the magnetic ordering and its relation to the ME coupling is provided by the ''magnetic phase shift'' mechanism involving $J_4$ and possibly $J_5$ exchange pathways.

\section{Conclusions}

We have investigated the magnetic ordering in the HT-IC phase of the FeTe$_2$O$_5$Br system by combining complementary neutron diffraction, nuclear quadrupolar and magnetic resonance techniques.
We find that due to the first-order transition from the paramagnetic phase all crystal symmetries are broken already in the HT-IC phase, which makes it, from the symmetry point of view, equivalent to the multiferroic LT-IC phase.
However, the ellipticity in the HT-IC phase is significantly reduced and the magnetic moments with sinusoidally modulated amplitudes align almost exactly along the $b$ axis.
Furthermore, the phase shifts between the magnetic AMOD waves, corresponding to the $J_4$ exchange interaction, converge towards $\Delta\psi_k\to0$ or $\pi$, which according to the {\em magnetic-phase-shift} ME coupling mechanism\cite{PregeljPRL} yields $P\propto M_{i}\cdot M_{j}\sin(\Delta\psi_k)$\,$\to$\,0, and could thus explain why the electric polarization vanishes in the HT-IC phase.\cite{PregeljPRL}
In addition, changes of the EFG at the Br$_2$ site in the LT-IC phase imply minute displacements of the Te$^{4+}$ ions, which must be associated with the emergent electric polarization.

\acknowledgments
We acknowledge the financial support of the Slovenian Research Agency (project J1-2118) and the Swiss National Science Foundation (project No.~200021-129899).
Neutron diffraction experiments were performed at SINQ, PSI, Villigen, Switzerland.

\appendix
\section{Calculations of NMR and NQR spectra}\label{appA}

Usually the NMR and NQR absorption lines are calculated for the two limiting cases.
In case of NMR experiment, Zeeman term in the nuclear spin Hamiltonian is typically taken as the dominant one, whereas electric quadrupole effects are considered as a perturbation.
On the other hand, for NQR, so-called Zeeman perturbed electric quadrupole Hamiltonian is assumed.
In the case of FeTe$_2$O$_5$Br, however, the NQR frequencies are comparable to the NMR ones and thus exclude the possibility to use simple perturbative approaches.
Therefore, our approach to calculate the resonance frequencies and their intensities is based on the exact diagonalization\cite{Bain} of the complete Hamiltonian for magnetic resonance of quadrupolar nuclei \cite{Slichter}
\begin{equation}
\mathcal{H}=\mathcal{H}_Z+\mathcal{H}_Q+\mathcal{H}_{\text{hyp}}+\mathcal{H}_{\text{dip}}.
\end{equation}
Here $\mathcal{H}_Z$ denotes the Zeeman term, $\mathcal{H}_Q$ the quadrupole interaction, and $\mathcal{H}_{\text{hyp}}$ and $\mathcal{H}_{\text{dip}}$ the influence of the hyperfine and dipolar fields, respectively.
The individual terms have the following form, 
\begin{equation}
\mathcal{H}_Z =-\gamma\hbar\,{\bf B_0}\cdot{\bf I},
\end{equation}
\begin{align}
\mathcal{H}_Q=\frac{eQ}{4I(2I-1)}[V_0(3I_z^2-I^2)+V_{+1}(I^-I_z+I_zI^-)\nn\\
+V_{-1}(I^+I_z+I_zI^+)+V_{+2}(I^-)^2+V_{-2}(I^+)^2],
\end{align}
\begin{equation}
\mathcal{H}_{\text{hyp}}=-\gamma\hbar\,{\bf \langle S\rangle}\cdot {\bf \hat {A}}\cdot{\bf I}, 
\end{equation}
\begin{equation}
\mathcal{H}_{\text{dip}}=-\gamma\hbar\,{\bf B_{\text{dip}}}\cdot{\bf I}.
\end{equation}
Here $\gamma$ denotes the nuclear gyromagnetic ratio, {\bf I}\,=\,($I_x$,\,$I_y$\,$I_z$) is the nuclear spin, $I^\pm$\,=\,$I_x$$\pm$$iI_y$, ${\bf \hat {A}}$ is the hyperfine coupling tensor, $V_0$\,=\,$V_{zz}$, $V_{\pm1}$\,=\,$V_{zx}$$\pm$$iV_{zy}$, $V_{\pm2}$\,=\,$\frac{1}{2}(V_{xx}$$-$$V_{yy})\pm iV_{xy}$ with $V_{ij}$ being the components of the EFG tensor, ${\bf \langle S\rangle}$ is the time averaged electron magnetic moment, and {\bf B}$_{\text{dip}}$ and {\bf B}$_0$ are the dipolar and the applied external magnetic fields, respectively.
In the long-range ordered magnetic states ${\bf \langle S\rangle}$ and {\bf B}$_{\text{dip}}$ are exactly determined by the LT- and HT-IC magnetic structures given in Ref.\,\onlinecite{PregeljPRL2} and Table\,\ref{tab1}, respectively, whereas in the paramagnetic phase magnetic moments are assumed to lie along {\bf B}$_0$ with amplitudes scaled by the magnetic susceptibility.\cite{PregeljPRB}
Exact diagonalization of the above Hamiltonian allows calculation of the nuclear spin eigenstates and thus enables to extract corresponding NMR/NQR transition frequencies for any kind of local (external or internal) magnetic field and for an arbitrary EFG tensor.
\begin{figure} [!]
\includegraphics[width=0.50\textwidth]{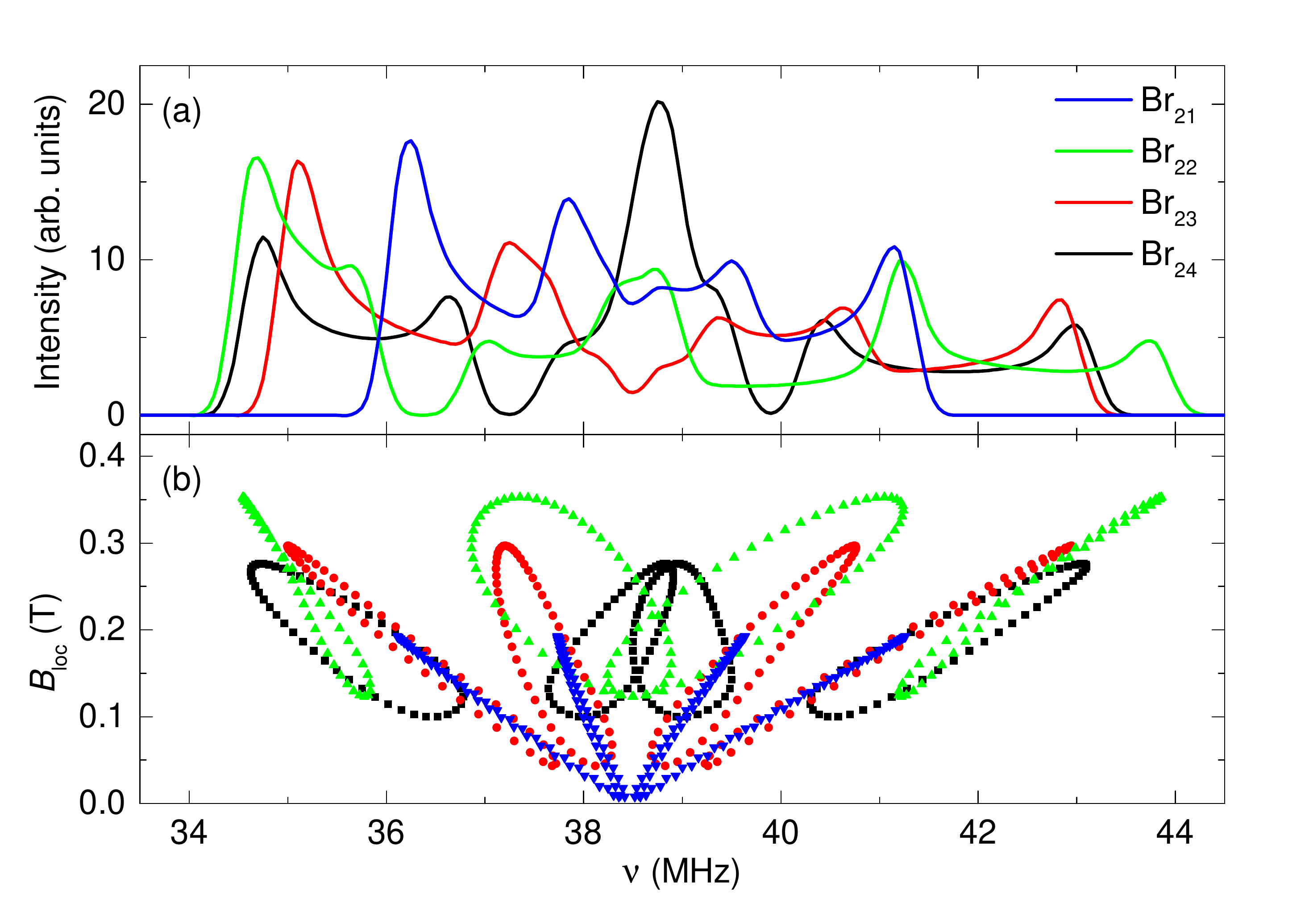}
\caption{(a) The simulated spectra for individual $^{81}$Br$_{2n}$ sites for a single domain. (b) Corresponding distributions of the amplitude of the local magnetic fields  {\bf B}$_{\text{loc}}$\,=\,$-\gamma h ({\bf \langle S\rangle}\cdot {\underline A}+{\bf B}_{\text{dip}})$ for the four contributing transitions between different nuclear spin states. We note that the intensity of the calculated spectra reflect the density distribution of the local fields, i.e., the peaks in the spectra coincide with the most common local fields.}
\label{figFieldsBr2}
\end{figure}
In addition, considering the orientation of the excitation/pick-up coil one can derive also the probabilities of the individual magnetic transitions and can thus estimate the intensities of the corresponding NMR/NQR absorption lines.
By rotation of the crystal system with respect to the laboratory system, the NMR/NQR spectrum, which for Br nucleus with $I$\,=\,3/2 in most general case consists of six absorption lines, can be calculated for any orientation of the crystal.

The $^{79,81}$Br NQR spectra in the IC long-range ordered magnetic phases were computed for each of the two magnetic domains (NMR/NQR experiments differentiate the two domains related by 2$_{1y}$ symmetry) by summing the four ($n$\,=\,1-4) spectral contributions for one crystal unit cell and than by summing the contributions of 100 consecutive cells along the IC direction ($b$ axis), as shown for a single domain in Fig.\,\ref{figFieldsBr2} for $^{81}$Br.
The complete spectra (Fig.\,\ref{figNQRsim}) were finally obtained as a sum of two contributions corresponding to the two equally populated domains. 
We note that for each crystallographicaly unique Br site (Br$_1$ or Br$_2$) electronic dipolar fields were calculated individually at each Br nucleus considered in the above summation (2$\times$4$\times$100) by assuming a sphere large enough ($\sim$40\,\AA) to ensure convergence.

\section{DFT calculations of EFG tensors}\label{appB}

The components of the EFG tensor were calculated {\it ab initio} within
the framework of the density-functional theory by applying the Wien97 
code,\cite{Blaha:1990} which adopts the 
full-potential linearized-augmented-plane-waves (FLAPW) 
method.\cite{Wimmer:1981}  The experimental data for the lattice parameters
and the atomic positions served to describe the input crystal structure at room temperature,\cite{Becker} 
whereas 
the muffin-tin radia were $2.1\>{\rm a.u.}$
for the Fe atoms, $1.98\>{\rm a.u.}$ for the Te atoms, $1.5\>{\rm a.u.}$ 
for the O atoms, and $2.68\>{\rm a.u.}$ for the Br atoms. 
The exchange-correlation effects were treated within the local-density
approximation (LDA).\cite{Perdew:1992-1}
The integration over the Brillouin zone (BZ) was discretized by summing up
333 {\bf k}-vectors in terms of the Gaussian method \cite{Fu:1983} with
the smearing parameter of $0.02\>{\rm Ry}$. 
The plane-wave-expansion  cut-off energy was set to $16\>{\rm Ry}$ and
the magnitude of the largest wave vector in the Fourier expansion of the 
charge density was $10\>{\rm a.u.}^{-1}$. 
The EFG-tensor components are defined as the second derivates of the Coulomb
potential at the particular nucleus. The Coulomb potential is obtained 
from the total charge density by Solving the Poisson's equation.
The calculation of the EFG is therefore straightforward once the non-spherical 
components of the charge density $\rho(r)$ are available as it is the case in 
the FLAPW method.

\end{document}